\documentstyle[preprint,aps]{revtex}

\begin{document}
\draft
\preprint{RU9552}
\title{Exact methods for Campi plots}
\author{K. C. Chase and A. Z. Mekjian}
\address{Department of Physics, Rutgers University \\
Piscataway, New Jersey 08855}
\date{\today}

\maketitle

\begin{abstract}
We introduce for canonical fragmentation models an exact method for
computing expectation values which exclude the largest cluster.
This method allows for the computation of the reduced multiplicity and
other quantities of interest introduced by Campi, and a
comparison shows that the percolation model and a recent canonical
model differ mostly only in small respects in these ensemble averages.
\end{abstract}
\pacs{25.70.Np, 21.65+f, 05.70.Ce, 05.70.Jk}

\narrowtext


Campi and Krivine~\cite{Campi1988,Campi1992}
introduced a method for distinguishing fragmentation models from one
another.  By comparing various expectation
values in which the largest cluster is excluded but the particle
number and fragment multiplicity are held fixed, they showed that
the percolation model has a distinctly different behavior than many
competing nuclear fragmentation models.

In this paper, we analyze another statistical weight we have been
using recently and show
that it shares many of the same properties as percolation theory, a
point already apparent from a consideration of critical
exponents~\cite{Chase1995}.
The method used in this paper is unusual in that it is an exact
computational method.  Monte Carlo sampling is avoided by exploiting
some properties of the partition functions, and as such there are no
statistical errors, an immense improvement over earlier
methods.


We begin by assuming that each fragmentation outcome happens with a
probability proportional to the Gibbs
weight~\cite{Mekjian1990,Mekjian1991,Mekjian1992,Cole1989}
\begin{equation}
W(\vec{n})
  = \prod_{k \ge 1} {x_{k}^{n_{k}} \over n_{k}!} \;,
\label{eq:weight}
\end{equation}
where $n_{k}$ is the number of fragments of size (or charge) $k$,
and $x_{k}$ is a parameter associated with $k$ sized fragments.
We then define the microcanonical partition function as
\begin{equation}
Z_{A}^{(m)}(\vec{x})
  = \sum_{\pi_{m}(A)} W(\vec{n}) \;,
\end{equation}
where $\pi_{m}(A)$ is the set of partitions of $A$
nucleons into $m$ fragments,
i.e. $\sum_{k} k n_{k} = A$, $\sum_{k} n_{k} = m$.
These partition functions satisfy the identity
\begin{equation}
{\partial Z_{A}^{(m)} \over \partial x_{k}}
  = Z_{A-k}^{(m-1)}(\vec{x}) \;,
\end{equation}
which allows for the computation of the partition functions
recursively from $\sum_{k} \langle n_{k} \rangle = m$, since
\begin{equation}
\langle n_{k} \rangle
  = x_{k} {\partial \over \partial x_{k}} \ln Z_{A}^{(m)}
  = x_{k} {Z_{A-k}^{(m-1)} \over Z_{A}^{(m)}} \;.
\end{equation}

Campi defined reduced moments as moments in which
the largest cluster is excluded from the measure, i.e.
\begin{equation}
M_{s}(\vec{n}) \equiv \sum_{k} k^{s} n_{k} - k_{\rm max}^{s}
\end{equation}
where $k_{\rm max}$ is the size of the largest cluster.
This suggested the definition of the reduced variance
$\gamma_{2}$~\cite{Campi1988,Campi1992} for a fragmentation event
should be
\begin{equation}
\gamma_{2}(\vec{n}) = {M_{2} M_{0} \over M_{1}^{2}} = (m-1)
  {\sum_{k} k^2 n_{k} - k_{\rm max}^{2} \over
    (A - k_{\rm max})^{2}} \;,
\end{equation}
Its expectation value can be computed by breaking events into classes
specified by $k_{\rm max}$, and summing the expectation values
over those classes with the appropriate weight, i.e.
\begin{equation}
\langle \gamma_{2} \rangle = (m-1)
  \sum_{k_{\rm max}} {\rm Pr}(k_{\rm max})
    {\sum_{k} k^{2} \langle n_{k} \rangle(k_{\rm max}) -
    k_{\rm max}^{2} \over (A-k_{\rm max})^{2}} \;,
\end{equation}
where ${\rm Pr}(k_{\rm max})$ is the probability of
$k_{\rm max}$ being the largest cluster size and
$\langle n_{k} \rangle(k_{\rm max})$ is the expectation
value of $n_{k}$ when $k_{\rm max}$ is fixed.

To compute expectation values in which the largest cluster size is
fixed we need to compute the partition function for such ensembles.
Clearly this partition function is given
by all the terms in the microcanonical partition function which have
$x_{k_{\rm max}}$ as the highest $x_{k}$ in the term.
Consider
\begin{eqnarray}
\Delta Z_{A}^{(m)}(k_{\rm max}) & \equiv &
     Z_{A}^{(m)}(x_{1}, \ldots, x_{k_{\rm max}}, 0, \ldots, 0)
     \nonumber \\
  && -Z_{A}^{(m)}(x_{1}, \ldots, x_{k_{\rm max}-1}, 0, \ldots, 0) \;.
\label{eq:deltaZ}
\end{eqnarray}
We see that $\Delta Z_{A}^{(m)}(k_{\rm max})$ is the partition
function for ensembles with fixed maximum cluster size
$k_{\rm max}$, since the first term collects all terms with
$x_{k_{\rm max}}$ or lower, and the second term eliminates those
terms which don't have an $x_{k_{\rm max}}$.  From this result
we can determine ${\rm Pr}(k_{\rm max})$ and
$\langle n_{k} \rangle(k_{\rm max})$, which are given by
\begin{eqnarray}
{\rm Pr}(k_{\rm max})
  & = &  {\Delta Z_{A}^{(m)}(k_{\rm max}) \over
          Z_{A}^{(m)}(x_{1}, \ldots, x_{A})} \\
\langle n_{k} \rangle(k_{\rm max})
  & = & \left\{
\begin{array}{ll}
  0 & k > k_{\rm max} \\
  {Z_{A-k}^{(m-1)}(x_{1}, \ldots, x_{k_{\rm max}}, 0, \ldots, 0) \over
   \Delta Z_{A}^{(m)}(k_{\rm max})} & k = k_{\rm max} \\
  {\Delta Z_{A-k}^{(m-1)}(k_{\rm max}) \over
   \Delta Z_{A}^{(m)}(k_{\rm max})} & k < k_{\rm max}
\end{array} \right.
\end{eqnarray}
This method is quite general and can be applied to other models.  For
example, equipartitioning models, which have weights given by
\begin{equation}
W(\vec{n}) = \prod_{k \ge 1} x_{k}^{n_{k}}
\end{equation}
can also be analyzed by this method with some minor modifications.
For example, $x_{k} = 1$ is the model used by Sobotka and
Moretto~\cite{Sobotka1985}.

With these identities there is sufficient information to compute
$\langle \gamma_{2} \rangle$ and other reduced moments for any
$x_{k}$.  We use $x_{k} = x/k^{\tau}$ for a variety of reasons
discussed elsewhere~\cite{Chase1994}.
Campi and Krivine~\cite{Campi1992} following Mekjian~\cite{Mekjian1990}
considered this model with $\tau = 1.0$ and showed
that its reduced variance and other related expectation values
had a distinctly different behavior than the percolation
model.  Plotting the expected reduced variance vs.
$(m-1)/A$, $\langle \gamma_{2} \rangle(m)$  has a single peak.
The location, height and width of this peak for the two models (and
other models they considered) are completely different,
suggesting the usefulness of this plot in distinguishing fragmentation
models.  The choice $\tau = 0$ was considered by Gross,
et.~al.~\cite{Gross1992a,Gross1992b,Jaqaman1992}, and a different
model was analyzed by Pan and Das Gupta~\cite{Pan1995}.

Since that time our interest has turned to the choice
$\tau = 2.5$ because of similarities with percolation theory and Bose
condensation.  Namely, the sudden appearance of an infinite cluster in
the infinite $A$ limit and the presence of condensation phenomena.  As
such, we have recomputed the Campi plots for
this model and have discovered that they
duplicate the percolation model results in many respects.
Figure~\ref{fig:gamma2kmax}(a) shows the results.
The height and location of the peak
of the reduced variance $\langle \gamma_{2} \rangle$
are the same in both models.   The only significant difference is the
width of the peak which is larger in the Gibbs model than
in the percolation model.
Plots of $\langle k_{\rm max} \rangle(m)$ vs. $(m-1)/A$ given
in Fig.~\ref{fig:gamma2kmax}(b) are also
very similar for both models, and the scaling behavior of the
position, width and height with changing $A$ also agree.

Another plot suggested by Campi~\cite{Campi1986} is to divide the
event space by
the maximum cluster size of the event
$k_{\rm max}$ and the reduced second moment $M_{2}$ and
plot the probability of the canonical model being at any particular
point on the graph.  This can be also be done exactly for Gibbs
models in a way completely analogous to the way given above.  Define
the partition function $Z_{A}(m_{2}; \vec{x})$ as the sum of the Gibbs
weight Eq.~(\ref{eq:weight}) over all partition vectors $\vec{n}$
which satisfy $\sum_{k} k n_{k} = A$, $\sum_{k} k^{2} n_{k} = m_{2}$.
This can be computed by the following recursion,
\begin{equation}
Z_{A}(m_{2}; \vec{x})
  = {1 \over A} \sum_{k} k x_{k} Z_{A-k}(m_{2}-k^{2}; \vec{x})
\end{equation}
with $Z_{0}(m_{2}; \vec{x}) = \delta_{m_{2},0}$.  If we define
$\Delta Z$ as in Eq.~(\ref{eq:deltaZ}), we find again the partition
function conditioned on $k_{\rm max}$ being fixed, which is
proportional to the probability of having an event with both $m_{2}$
and $k_{\rm max}$.  Figure~\ref{fig:campiplots} plots this
probability profile as a contour plot, which reveals the events are
centered on a particular region in this phase space.  The slopes of the
edges of this region are related to the critical exponents
according to Campi~\cite{Campi1986}.


Clearly there are differences between percolation theory and a
Gibbs model, but the differences are not as large as
originally suggested by early computations.  Indeed the reduced
variance might not reliably distingush percolation from a simple
Gibbs model.  A different method is needed to distingush these
models.  However, the idea of excluding the largest cluster from the
ensemble averages is a standard procedure in percolation
theory~\cite{Stauffer1992}, and this new technique for doing that
analytically in the Gibbs models shows a particular advantage of
these models over percolation models, which we hope will encourage
further interest in them.

This work supported in part by the National Science Foundation
Grant No. NSFPHY 92-12016.

\begin{figure}
\caption{Expected reduced multiplicity $\langle \gamma_{2} \rangle(m)$
(a) and largest cluster size $k_{\rm max}/A$
(b) vs. $(m-1)/A$ for $\tau = 1.0, 2.5$ and the percolation model at $A=125$.}
\label{fig:gamma2kmax}
\end{figure}

\begin{figure}
\caption{Campi probability contour plot for $Z=79$, $\tau = 2.5$ at
the critical point $x = x_{c}$.
The axes are logarithmic, with the largest cluster size
on the $y$-axis, and the ratio of the reduced moments $M_{2}/M_{1}$ on
the $x$-axis.  The central rings are higher in probability than the
outer rings.}
\label{fig:campiplots}
\end{figure}

\end{document}